\documentclass[sn-basic]{sn-jnl}

\usepackage{graphicx}%
\usepackage{multirow}%
\usepackage{amsmath,amssymb,amsfonts}%
\usepackage{amsthm}%
\usepackage{mathrsfs}%
\usepackage[title]{appendix}%
\usepackage{xcolor}%
\usepackage{textcomp}%
\usepackage{manyfoot}%
\usepackage{booktabs}%
\usepackage{algorithm}%
\usepackage{algorithmicx}%
\usepackage{algpseudocode}%
\usepackage{listings}%
\usepackage{siunitx}

\newgeometry{paper=letterpaper,top=1in,bottom=1in,left=1in,right=1in}
\usepackage{mathtools}
\usepackage{bm}
\usepackage{comment}

\newcommand{\mbf}[1]{\mathbf{#1}}

\newcommand{\tE}{\mathbf{E}}

  \newcommand{\bX}{\mbf{X}}

  \newcommand{\grad}{\boldsymbol{\nabla}} 
  \newcommand{\divg}{\grad\cdot} 

  \newcommand{\bsigma}{\boldsymbol{\sigma}}

\def\XXint#1#2#3{{\setbox0=\hbox{$#1{#2#3}{\int}$}
     \vcenter{\hbox{$#2#3$}}\kern-.5\wd0}}

\newcommand{\MDGrevise}[1]{\textcolor{black}{#1}}

\newcommand{\MEHrevise}[1]{\textcolor{black}{#1}}
\newcommand{\Formsax}{FFoRM-sSAXS}

\usepackage[normalem]{ulem}

\theoremstyle{thmstyleone}%
\theoremstyle{thmstyletwo}%

\theoremstyle{thmstylethree}%

\begin{document}

\title[]{Scattering-Informed Microstructure Prediction during Lagrangian Evolution (SIMPLE) -- A data-driven framework for modeling complex fluids in flow}


\author[1]{Charles D. Young}
\author[2]{Patrick T. Corona}
\author[2]{Anukta Datta}
\author*[2]{Matthew E. Helgeson}\email{helgeson@ucsb.edu}
\author*[1]{Michael D. Graham}\email{mdgraham@wisc.edu}

\affil[1]{\orgdiv{Department of Chemical and Biological Engineering}, \orgname{University of Wisconsin-Madison}, \orgaddress{\city{Madison},  \state{Wisconsin}, \postcode{53706}, \country{USA}}}

\affil[2]{\orgdiv{Department of Chemical Engineering}, \orgaddress{\orgname{University of California}, \city{Santa Barbara}, \state{California}, \postcode{93106-5080}, \country{USA}}}


\abstract{An overarching challenge in rheology is to develop constitutive models for complex fluids for which we lack accurate first principles theory. \MDGrevise{A further challenge is that most experiments probing dynamical structure and rheology do so only in very simple flow fields that are not characteristic of the complex deformation histories experienced by material in a processing application. } A recently-developed experimental methodology holds potential to overcome this challenge by incorporating a fluidic four-roll mill (FFoRM) into scanning small-angle x-ray scattering instrumentation (sSAXS) [Corona, P. T. et al. \emph{Sci.~Rep.} \textbf{8}, 15559 (2018);
    Corona, P. T. et al. \emph{Phys.~Rev.~Mater} \textbf{6}, 045603 (2022)]
  to rapidly generate large data sets of scattering intensity for complex fluids along diverse Lagrangian flow histories. \MDGrevise{To exploit this uniquely rich FFoRM-sSAXS data, we propose a machine learning framework,  \emph{Scattering-Informed Microstructure Prediction under Lagrangian Evolution} (SIMPLE), which uses FFoRM-sSAXS data to learn an evolution equation for the scattering intensity and an associated tensorial differential constitutive equation for the stress.}  The framework incorporates material frame indifference and invariance to arbitrary rotations by data preprocessing. We use autoencoders to find an efficient reduced order model for the scattering intensity and neural network ordinary differential equations to predict the time evolution of the model coordinates. The framework is validated on a synthetic FFoRM-sSAXS data set for a dilute rigid rod suspension. The model accurately predicts microstructural evolution and rheology for flows that differ significantly from those used in training. SIMPLE is compatible with but does not require material-specific constraints or assumptions. 
}

\keywords{rheo-scattering, constitutive modeling, physically-informed machine learning, structure-property relationships}



\maketitle

\section{\label{sec:Intro}Introduction}

There is a critical need to develop rational design approaches for the flow-processing of microstructured fluids and soft materials. In particular, researchers have recently recognized the important role that the \textit{complexity} of processing flows plays in the development of these materials \citep{diao2013solution}. Unlike prototypical flows (simple shearing or elongation) in which the type and rate of deformation are uniform everywhere in the flow, most industrially-relevant processing flows involve complex non-uniform flow fields, in which fluid elements experience significant variations in deformation type and rate as they move along the process. For example, in film casting processes, the fluid experiences a transition from low-rate shearing to extension and then to high-rate shearing as it passes across a coating blade \citep{diao2013solution}. Such complex flows can significantly influence the final structure and performance of thin films obtained from these and other processes.  

Although researchers have begun to leverage this understanding to optimize flow processes toward desired structures, most approaches remain primitive -- involving empirical searching of parameters until optimal structure or performance is achieved. This is because process-structure relationships are often developed from \textit{ex situ} measurements, and correlated to independently specified process parameters (\textit{e.g.}, blade height, substrate velocity, etc.). However, this approach is process-specific, and as such is unable to anticipate new flow processes that could dramatically open the range or optimization of processed structures available to a material. This has imposed a significant bottleneck in development of flow-processed materials in important technologies such as printed electronics \citep{qu2016flow,patel2017multiscale}, photonics \citep{demirors2022three}, and solar cells \citep{du2022flow}.

An alternative and much more powerful approach would be founded on detailed modeling of flow-microstructure interactions to predict what a material of interest might do in \textit{any possible process flow}. In such a framework, one would seek to parameterize not a process, but a \textit{flow history} of a material element as it moves through the process. The natural setting for describing such flow histories is the Lagrangian frame, which maps an arbitrary flow, (\textit{i.e.}, an Eulerian spatial distribution of velocities $\bm{v}(\bm{x})$) onto a set of time-dependent trajectories involving variations in the velocity gradient, $\grad \bm{v}(\mathcal{T})$, where $\mathcal{T}$ is elapsed time along a streamline. In this frame, both the flow itself as well as the structural evolution of the material can be uniquely determined through time-variations in the components of $\bm{\nabla v}$, thereby collapsing all possible processing flows to a set of Lagrangian excursions in these low-dimensional variables.  To be useful, this approach would require: (1) measurements of microstructure and velocity gradient along these Lagrangian trajectories, (2) a modeling framework that could take this information to generate an evolution equation for the microstructure and ultimately the stress tensor. 

\begin{figure}[h]
    \centering
\includegraphics[width=0.95\textwidth]{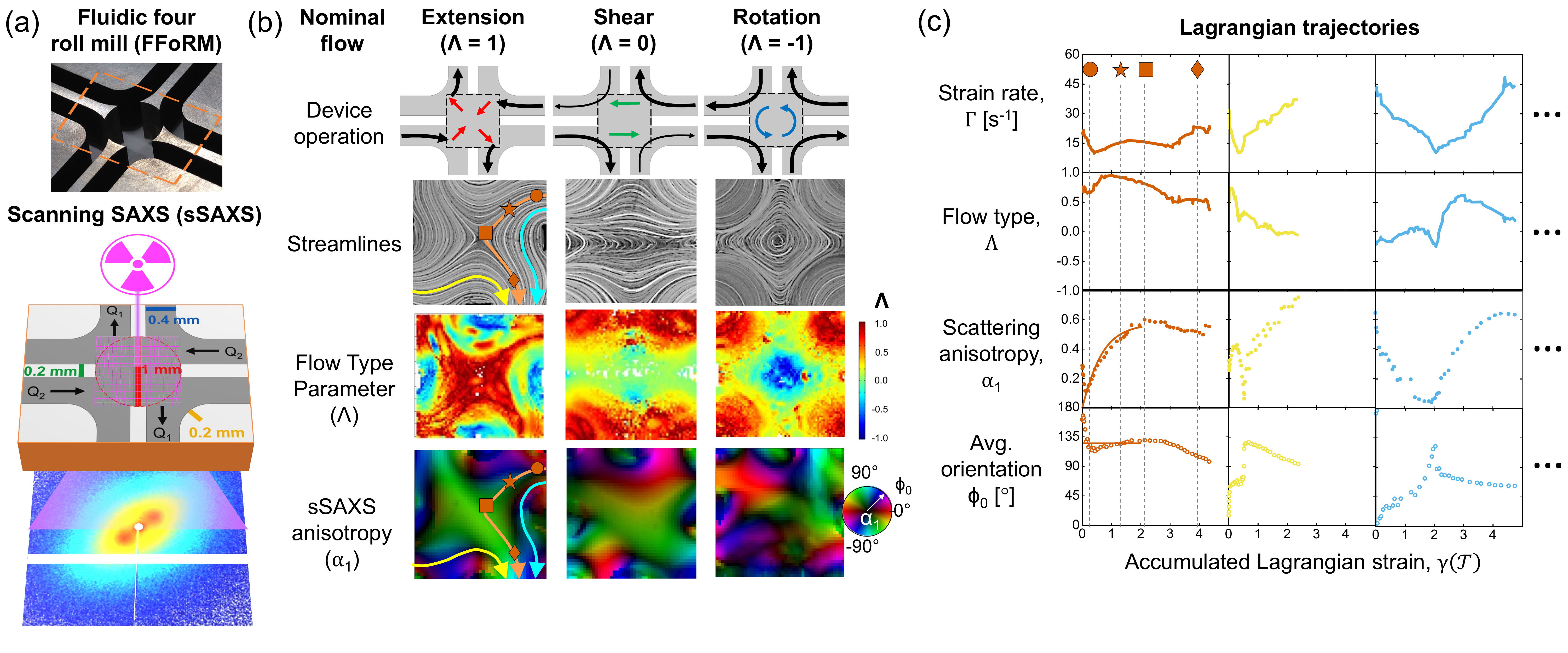}
    \caption{\textbf{Overview of the FFoRM-sSAXS method (adapted from \citep{corona_fingerprinting_2022}).} (a) Photograph of the FFoRM geometry and schematic with dimensions illustrating the scanning sSAXS measurement. (b) Operation of the FFoRMin three distinct quasi-2D flows, controlled by the direction and magnitude of imposed flow rates (black arrows), and the resulting multi-modal Eulerian FFoRM-sSAXS spatial data. (c) Conversion of the Eulerian data along streamlines in (b) produces corresponding Lagrangian trajectories of flow and structural data.}
\vspace{-10pt}
    \label{fig:Fig2}
\end{figure}

Newly developed spatially resolved scattering measurements in complex flows hold exceptional promise toward the first of these requirements. In particular, Helgeson and coworkers recently adapted a ``fluidic four roll mill'' (FFoRM), a version of Taylor's original four roll mill \citep{taylor1934formation} due to Mueller and co-workers \citep{lee2007microfluidic}, for beamline scattering measurements including high-resolution scanning small angle X-ray scattering (sSAXS) \citep{corona_probing_2018,corona_fingerprinting_2022}. The velocity field in these flows can be quantified by particle tracking velocimetry (PTV). Unlike other fluidic devices for \textit{in situ} scattering \citep{eberle2012flow,lopez2015microfluidic,lopez2018microfluidic,lutz2015micellar,bent2003neutron, donina2021lamellar,silva2015nematic,poulos2016microfluidic,martin2016microfluidic,lutz2016scanning}, the FFoRM can achieve a range of quasi-2D stagnation point flows for various rheologically complex fluids, enabling a wide array of flow histories (and microstructures in them) to be mapped from the same device (Fig. \ref{fig:Fig2}). This FFoRM-sSAXS method is therefore ideal for generating the large, diverse flow-microstructure data sets required to achieve our aims. Helgeson and collaborators have already generated large data sets \citep{corona_fingerprinting_2022}, and have developed state-of-the-art scattering models to draw physical insights from them \citep{corona_bayesian_2021,corona2023testing,corona2020probing}. For example, FFoRM-sSAXS on rod-like nanoparticle suspensions was used to distinguish the influence of interparticle interactions on flow-induced alignment \citep{corona_fingerprinting_2022}. Furthermore, although this has not yet been done, in principle it is possible, with a variational data assimilation method \citep{Asch:2016}, to use the measured velocity fields to infer the particle contribution to the stress tensor, as elaborated in Sec. \ref{sec:methodology}.

With regard to the second requirement, one approach would be to start with classical physics-based approaches for development of constitutive models. There are some clear successes for this approach, for example in highly entangled polymers \citep{bird1987dynamics,doi1980rheology,sato2020review} and concentrated hard sphere Brownian particle suspensions \citep{morris2009review,morris2020shear,morris2020toward}. 
Despite these prominent successes, first principles rheological theory has failed to keep up with the ever-expanding set of materials whose flow processing is crucial for emerging technologies. This is because the development of microconstitutive models is slow, requiring iterative testing between theory and experiment to propose and validate physical theories for a material's structural dynamics. It is also biasing, in that rheological experiments are limited to situations in which the fluid is subjected to a homogeneous deformation type and rate. As a result, while a model may perform accurately in rheometric flows, it may do quite poorly in more general flows. 
A prototypical example is non-dilute suspensions of elongated (rod-like, ellipsoidal, etc.) Brownian particles. Although an accurate \textit{dilute} theory has been known for some time \citep{bird1987dynamics}, a successful \textit{non-dilute} theory is lacking due to the complicated coupling between orientational dynamics and many-body interactions. Dynamical theories \citep{doi1980rheology,doiedwards,dhont2003viscoelasticity} involving assumptions of mean-field interactions between particles and infinite aspect ratios to treat the many-body problem fail to quantitatively capture the features of flow-induced alignment and shear thinning even for steady state homogeneous shear flows in the semi-dilute regime \citep{lang2019microstructural}. Even when one resolves particle orientations exactly, theories for the rheology of these systems \citep{batchelor1970stress,hinch1972effect} provide accurate predictions only for linear, \textit{homogeneous} flows, with extensions to complex flows of the type we wish to address an active area of research \citep{ferec2022macroscopic}. 

Given the incredibly rich data available from the \Formsax \ experiment, data-driven modeling provides an alternate approach that could dramatically accelerate predictive design of flow-processed materials. In Newtonian fluid mechanics, data assimilation and machine learning approaches have had significant impact \citep{Brunton.2019.10.1146/annurev-fluid-010719-060214}. In seminal work, Ling et al.~\citep{Ling:2016fz} used turbulence simulations, neural networks (NNs), and model representations constrained by Galilean invariance to compute expressions for the Reynolds stress tensor (i.e.~a closure model) to facilitate Reynolds-averaged Navier-Stokes simulations of turbulence. Such advances are enabled by the large data sets that can be generated either by direct numerical simulation (DNS) or spatiotemporally resolved velocimetry experiments for Newtonian flows.

In rheology and non-Newtonian fluid mechanics, comparatively few applications of machine learning have been reported. Some have used rheometric data to generate parameters for given constitutive relations  \citep{Mahmoudabadbozchelou:2021hk, Mahmoudabadbozchelou:2021bf, marino2022automated, saadat_data-driven_2022}. Others have used a generalized upper convected Maxwell model with a nonlinear frame indifferent term in the form of a neural network, again trained on rheometric data \citep{Lennon.2022.10.48550/arxiv.2210.04431}. Related work involved learning evolution equations for a set of conformation tensors as conformational descriptors \citep{lei_machine-learning-based_2020}. Thakur et al.~\citep{Thakur.2022.10.48550/arxiv.2209.06972} described a physics-informed deep learning approach that uses a given velocity field to select a best-fit constitutive model and parameters (from a set of standard models). Despite these advances, the fields of rheology and non-Newtonian fluid mechanics have yet to embrace, exploit and advance data-driven methods for modeling and data assimilation. 


Several data-driven methods have been developed for efficient, scalable molecular dynamics simulations with \textit{ab initio} accuracy from atomistic coordinates of a DFT simulation \citep{behler2007generalized, chmiela2017machine, zhang_deep_2018, zhang_deepcg_2018, shireen_machine_2022}. These approaches essentially generate an accurate and generalizeable empirical force field which accounts for molecular symmetries, such as translation and rotation. So far these methods have primarily been applied to small molecules, but in principle they could be used for macromolecular soft matter systems. However, they do not perform dimension reduction, and so may not be suitable to spatially extended domains and long time scales required in soft matter simulations. Lei et al. \citep{lei_machine-learning-based_2020} proposed a model designed for non-Newtonian fluids which discovers coarse-grained frame-invariant configuration tensors based on the Fokker-Planck equation. The model discovers an appropriate coarse-grained description based on molecular coordinates, such that it can be extended to non-linear chain architectures \citep{fang_deepn2_2022}. However, the model does not explicitly account for intra- or inter-molecular hydrodynamic interactions, other than those arising from the simulation training data. An alternative coarse-graining approach is the Mori-Zwanzig formalism, in which the challenge is to learn the memory kernel for a generalized Langevin equation describing a stochastic observable. Several data-driven models have implemented this approach for the dynamics of small molecules and dilute polymer solutions \citep{grogan2020data, ma_transfer_2021, she2022data, russo_machine_2022}.

The above approaches rely on molecular simulation data, which is often computationally prohibitive to collect for novel materials of interest, particularly when screening a library of materials. While computational approaches are being developed to accelerate this process \citep{jackson_recent_2019}, it is often more practical to use experimental data. These data often come in the form of two-dimensional (2D) images or scattering patterns, which has motivated the use of convolutional NN (CNN) autoencoders (AEs). These have been use to featurize microstructural images and classify the resulting latent space, as shown for synthetic data \citep{lubbers2017inferring}, electronic microscopy of metals \citep{ling2017building}, and x-ray photon correlation spectroscopy of colloidal glass relaxation after cessation of shear flow \citep{horwath2022elucidation}. Beltran-Villegas et al. \citep{beltran2019computational} applied genetic algorithms and molecular simulations to recover the real-space structure of self-assembled amphiphilic polymer solutions from small-angle scattering patterns. Corona et al. \citep{corona_fingerprinting_2022} used small-angle scattering to fingerprint the deformation of rod suspension under mixed shear and rotational flows in a microfluidic four-roll mill \citep{corona_probing_2018}. Chen and coworkers have made substantial progress in analyzing scattering data for fingerprinting nonequilibrium dynamics \citep{shen2022fingerprinting}, inferring the potential of mean force \citep{chang2022machine}, and inversion to the microscopic distribution function \citep{huang2019orientational, huang2021exact}.

Beyond developing physical insight from microstructural measurements, it remains a significant challenge to determine to bulk mechanical response without training labels. Corona et al. \citep{corona_bayesian_2021} used Bayesian estimation to infer the orientation probability distribution function (OPDF) of a semidilute rigid rod suspension from small-angle scattering measurements. The OPDF was then used to determine the particle contribution to the stress. However, this approach is difficult to generalize to deformable materials in which the underlying microscopic distribution function is not on the surface of a sphere.

\begin{figure*}
    \centering
    \includegraphics[width=0.9\textwidth]{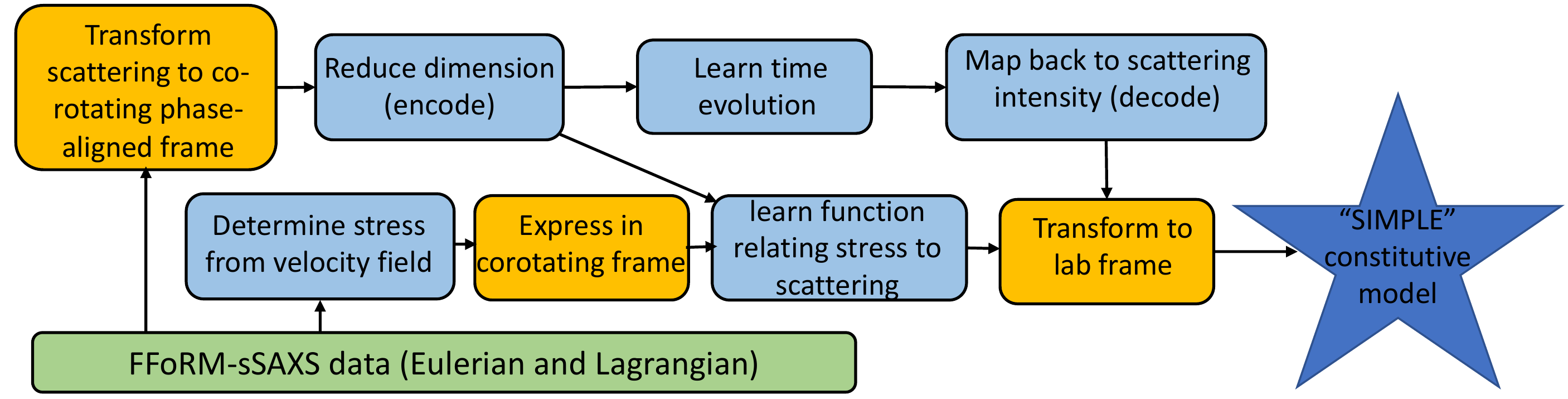}
    \caption{SIMPLE constitutive equation learning framework for FFoRM-sSAXS data. Scattering intensity and velocity measurements are first collected and transformed to the Lagrangian frame (green box). Large data sets enabled by FFoRM-sSAXS are used to train data-driven operations (blue boxes) of dimension reduction and time evolution.  Orange boxes are known operations accounting for frame indifference, phase alignment, and material symmetries.}
    \label{fig:modeling}
\end{figure*}

In the present work, we introduce a framework, which we call ``Scattering-Informed Microstructure Prediction during Lagrangian Evolution (SIMPLE)'', to leverage the large, high-dimensional data sets produced by FFoRM-sSAXS to develop a completely new capability --- a data-driven workflow to rapidly learn a general frame-indifferent constitutive model that can then be used to simulate that fluid in an arbitrary 2D flow.  The overall structure of SIMPLE is illustrated in Figure \ref{fig:modeling}. This paper will focus on development of the computational aspects of the data-driven constitutive modeling using synthetic scattering data determined from Brownian dynamics simulations of a dilute suspension of Brownian rigid rods \citep{Graham.2018.https://doi.org/10.1017/9781139175876} in a FFoRM flow field generated with the open source CFD code OpenFOAM. Future work will address the data-assimilation problem for inferring stress from velocity data and the development of models from true experimental scattering data for complex fluid systems. 

The remainder of the paper introduces the SIMPLE methodology. In Sec. \ref{sec:methodology}, we describe the framework and the data set we use to evaluate the quantitative and qualitative accuracy of the microstructural and rheological predictions. In Sec. \ref{subsec:training}, we quantify the performance of the model on interpolated flow protocols, which are not used in training but are within the training data range. Then in Sec.  \ref{subsec:extrapolation}, we test the trained model on extrapolation to unseen flow protocols. Finally, in Sec. \ref{sec:rheology} we demonstrate the ability of NNs to learn the particle contribution to the stress from the scattering intensity.

\section{Methodology \label{sec:methodology}}

\newcommand{\lagtime}{\mathcal{T}}
\noindent The aim of the data-driven SIMPLE framework is to use FFoRM-sSAXS data to develop microconstitutive models for complex fluids. This framework, shown in Figure \ref{fig:modeling}, will infer stress information from velocity data (we do not address this issue here), extract the essential microstructural degrees of freedom from scattering data, and determine evolution equations for the microstructure and stress using velocity gradient and scattering data along streamlines. 

\subsection{Initial data analysis}
 Data from the FFoRM-sSAXS experiments will be steady-state two-dimensional velocity fields $\bm{v}(\bm{x})$ and spatially-resolved scattering patterns $I(\bm{q},\bm{x})$. From each velocity field, a large number of Lagrangian particle trajectories will be generated by choosing points $\bX_0$ in the domain and integrating $d\bX/d\lagtime=\bm{v}(\bX(\lagtime))$ to yield trajectories $\bX(\bX_0,\lagtime)$.  Along each trajectory the rate of strain $\tE(\bX(\bX_0,\lagtime))$ and vorticity tensor $\bm{\Omega}(\bX(\bX_0,\lagtime))$ are computed. Relatedly, we can determine the scattering pattern along each trajectory, $I(\bm{q},\bX(\bX_0(\lagtime)))$.  In general we will suppress the dependence of Lagrangian quantities on their position $\bX(\bX_0,\lagtime)$. Recall that in the present work, we use synthetic data from Brownian dynamics simulations on trajectories in a flow field determined by CFD as detailed in Sec. \ref{subsec:data}, rather than true experimental data.


\subsection{Accounting for material frame indifference and rotational equivariance}

Even for a purely data-driven framework, key symmetries that data obey can be built into the reduced representation. For constitutive models, the key symmetry is material frame indifference, which reflects the physical idea that rigid rotations should not induce any relative motion within the microstructure. In the present work we enforce this by working in a reference frame corotating with the fluid element. The importance of including this physical constraint cannot be overstated -- indeed a widespread current theme in scientific ML is building symmetries into representations and architectures; otherwise neural networks require vast amounts of data to learn them, and even then only do so approximately.

\begin{figure*}
    \centering
    \includegraphics[width=1.0\textwidth]{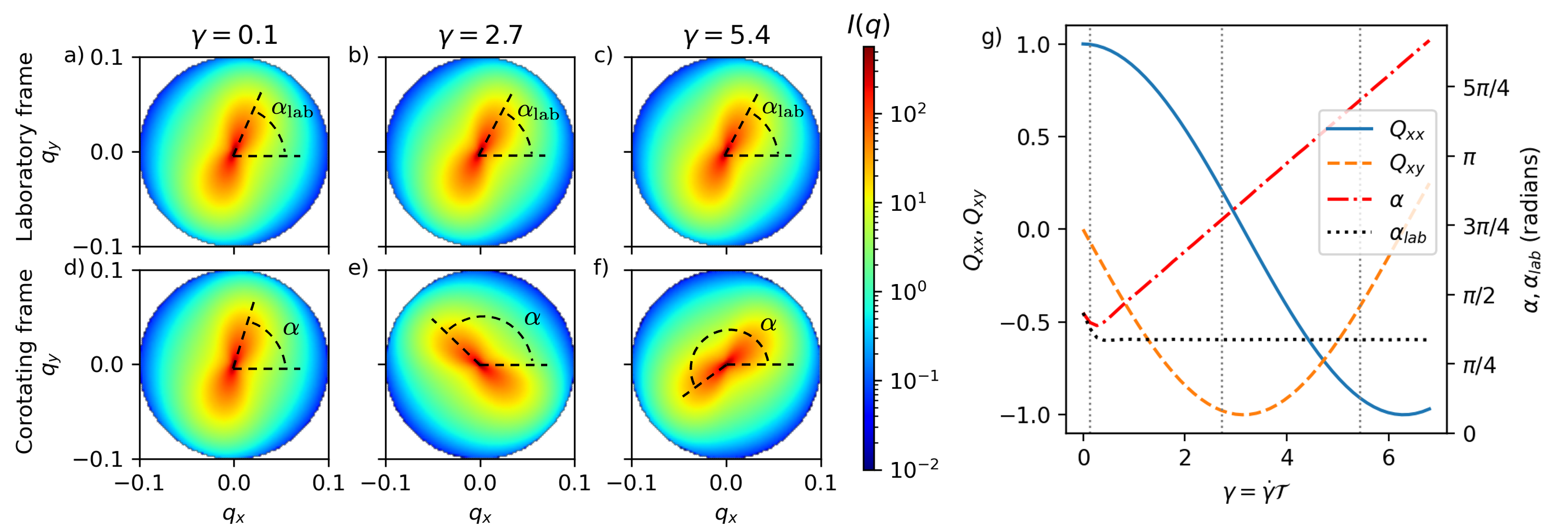}
    \caption{Pre-processing of startup steady shear flow ($\textrm{Pe} = 13.6$) data before data-driven operations. The lab frame data (a-c) are first transformed to the co-rotating frame (d-f) using the rotation matrix $\textbf{Q}$ determined from the vorticity tensor $\bm{\Omega}$. The data are then aligned by the phase of the first azimuthal Fourier mode of the data. g) Time evolution of the relevant rotation matrix components and the corotating frame phase $\alpha$. For comparison, we include the time dependence of the laboratory frame phase $\alpha_{lab}$. The dashed vertical lines correspond to the snapshot times.}
    \label{fig:corot_phase}
\end{figure*}

To achieve this for scattering data, we compute the unitary rotation matrix \textbf{Q} which relates the rotating frame basis vectors $\bm{e}_i^R$ to the stationary laboratory frame basis vectors $\bm{e}_i$ via $\bm{e}_i = \textbf{Q} \cdot \bm{e}_i^R$. The rotation of the reference frame is determined from the vorticity of the flow $\bm{\Omega} = (\bm{\nabla} \bm{v} - \bm{\nabla} \bm{v}^T)/2$ as 
\begin{equation}
    \frac{d\textbf{Q}}{d\mathcal{T}} = \bm{\Omega}^T \cdot \textbf{Q}
    \label{eqn:vort_evo}
\end{equation}
From the rotation matrix, we compute a rotation angle $\theta_{\bm{\Omega}} = \textrm{atan2} (\textrm{Q}_{xy},\textrm{Q}_{xx})$. This is used to perform a change of basis between the laboratory frame observation of the scattering intensity $I(\bm{q})$ and the co-rotating frame $I_{\bm{\Omega}}(\bm{q}) = \mathcal{R}(I(\bm{q}),\theta_{\bm{\Omega}})$, where $\mathcal{R}(\cdot,\theta)$ is the rotation operation. In the current study, we consider the flow to be 2D in the $xy$ plane and the scattering intensity to be observed as the 2D projection $I(q_x,q_y)$ on a uniform grid. The range and number of wavevectors in both directions is $q \in [0,0.1]$ \r{A}$^{-1}$ and $n_q = 100$. This is lower resolution than many scattering experiments (typical resolutions are $n_q = 128$ for neutron scattering, $n_q = 1475$ for the PILATUS 2M detector at the Paul Scherrer Institute used in \cite{corona_fingerprinting_2022}, and $n_q = 1024$ for the LiX beamline at NSLS-II (Brookhaven National Laboratory)). The data are radially truncated to $r = \sqrt{q_x^2 + q_y^2} < 0.1$ \r{A}$^{-1}$ so the image can be rotated without loss of information. An example of this rotation and the rotation matrix in steady shear flow is shown in Fig. \ref{fig:corot_phase}.

Material frame indifference is a consequence of physical assumptions about the evolution of microstructure.  Another symmetry that applies more generally is simple equivariance under rotations.  For example, the scattering pattern for a material under extensional flow along the $x$-direction should  be a rotation of the same pattern for the fluid stretched along the $y$-direction. A brute force solution to approximate this symmetry is to augment the training data with rotated representations, but this means dramatically increasing the size of the data set while still only approximately satisfying the symmetry relation. While modern neural network architectures are making progress towards satisfying these conditions by construction \citep{winter2022unsupervised}, it is often simpler and more effective to apply these symmetry groups to the data before further processing \citep{ling2016machine, linot2020deep, zeng2021symmetry}. Following the latter approaches, we further preprocess the scattering images by computing a rotational phase angle and then rotating the images to the the same phase. We compute the radially averaged azimuthal dependence of the scattering intensity where $\Theta = \textrm{atan2}(q_y,q_x)$ \citep{walker1996sans, bunk_multimodal_2009}.
\begin{equation}
    I_{\bm{\Omega}}(m_r, n_\Theta) = \frac{1}{A_{m_r,n_\Theta}} \sum_{r \in R(m_r), \Theta \in \theta(n_\Theta)} I_{\bm{\Omega}}(r,\Theta).
\end{equation}
$A$ is the number of pixels in a segment of radial range $R$ and azimuthal range $\theta$, and $m_r$ and $n_\Theta$ are the radial and azimuthal indices respectively. We use $N_\Theta = 128$ azimuthal segments and a radial range $r \in [0.03,0.05]$ \r{A}$^{-1}$. The phase is computed from the first mode of discrete Fourier transform of the azimuthal average $z = \mathcal{F}(I_{\bm{\Omega}}(n_\Theta))$ as $\alpha = \textrm{atan2}(\textrm{Im}[z_1],\textrm{Re}[z_1])$.  The phase aligned data are generated by the rotation $I_{\bm{\Omega},\alpha} = \mathcal{R}(I_{\bm{\Omega}},\alpha)$. Thus, the rotational evolution is captured in the co-rotating frame phase $\alpha$, and the deformation pattern is captured by the phase-aligned scattering intensity. The corotating phase under steady shear flow is compared to the laboratory frame phase in Fig. \ref{fig:corot_phase}.

\subsection{Dimension reduction}
The machine learning task is now to determine the evolution of the scattering pattern and phase given an input deformation. However, the data are still high-dimensional ($I(\bm{q}) \in \mathbb{R}^d, d = n_q^2 = 10^4$, and up to $d \approx 10^6$ for the experimental data collected on the beamlines listed above), which would incur significant computational expense in training NN time evolution models and deploying the model for structure prediction. \MDGrevise{Furthermore, we do not expect $n_q^2$ dimensions to be necessary for representation of the microstructure of the material \citep{lubbers2017inferring}.} We therefore seek a reduced order representation of the data to overcome these issues. We follow recent methods which use undercomplete autoencoders (AEs) to discover invariant manifolds on which data lie in chaotic and turbulent dynamical systems \citep{linot2020deep, linot2023dynamics}.

\begin{figure*}
    \centering
    \includegraphics[width=1.0\textwidth]{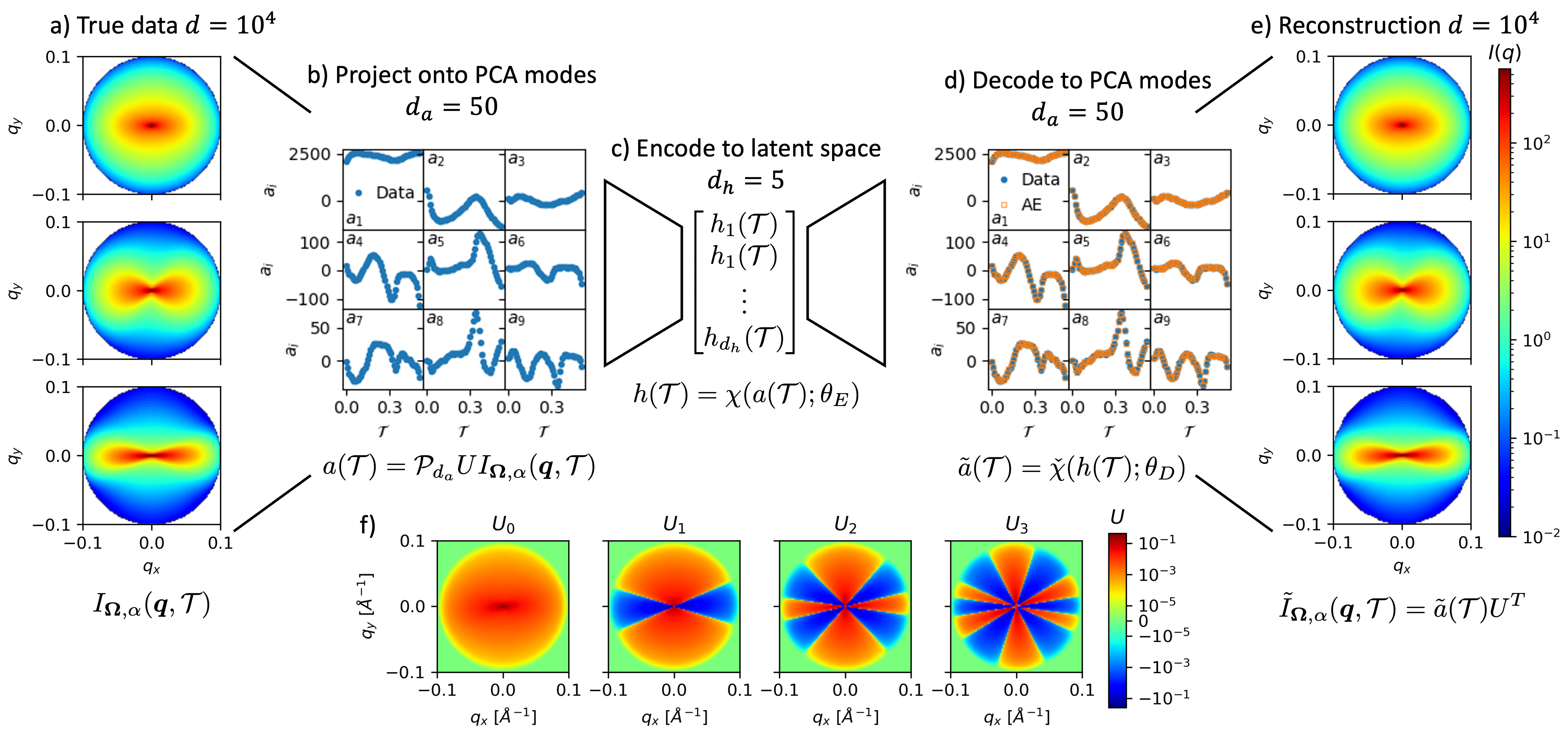}
    \caption{Dimension reduction of the phase-aligned data $I_{\bm{\Omega},\alpha}(\bm{q},\mathcal{T})$ for an FFoRM streamline from the test data (same as in Fig. \ref{fig:fform_viz}) (a) Snapshots of the data at $\lagtime = 0.01,0.24,0.52$.  Each snapshot is projected onto the leading $d_a = 50$ PCA modes $U$ (b) and then encoded to the latent space with dimension $d_h = 5$ (c) using a deep autoencoder. The PCA modes are reconstructed from the latent space using the decoder (d), and finally the scattering intensity is reconstructed (e). The first 4 PCA modes are visualized in (f).}
    \label{fig:autoencoder}
\end{figure*}

Before passing the data to the autoencoder, we use principal components analysis (PCA) \MDGrevise{to perform an initial coarse dimension reduction step}. PCA is a common linear dimension reduction technique by which the data are transformed to an orthogonal basis \MDGrevise{defined by the eigenvectors of the data covariance,} and truncated to retain directions in which the variance (``energy") is large. There are two notable advantages to performing PCA before the autoencoder. First, the resulting input to the AE is lower dimensional, reducing the expense of training the NN. Second, the ordering of PCA modes places \MDGrevise{lower-variance} fluctuations in the truncated trailing modes, which reduces noise resulting from experimental uncertainty or sampling limitations. We compute the PCA basis vectors by singular value decomposition of $M$ snapshots of the scattering intensity, $I(\bm{q},\lagtime) \in \mathbb{R}^{d \times M}$:
\begin{equation}
    I_{\bm{\Omega},\alpha}(\bm{q},\lagtime) = U \Sigma V^T
\end{equation}
where $U \in \mathbb{R}^{d \times d}$ is the square orthogonal matrix of PCA basis vectors, $\Sigma \in \mathbb{R}^{d \times M}$ is a diagonal matrix of singular values, and $V \in \mathbb{R}^{M \times M}$ is the orthogonal matrix of right singular vectors. Dimension reduction is performed by projecting the data onto the leading $d_a$ PCA basis vectors as $a = \mathcal{P}_{d_a} U I_{\bm{\Omega},\alpha}(\bm{q})$. The original data are reconstructed by $\tilde{I}_{\bm{\Omega},\alpha}(\bm{q}) = aU^T$. We set a criteria to retain 99 \% of the energy, which reduces the dimension to $d_a = 50$. The PCA modes of the data $a(\lagtime) \in \mathbb{R}^{d_a \times M}$ are then input to a deep autoencoder, which encodes the dimension to a latent space representation $h = \chi(a;\theta_E) \in \mathbb{R}^{d_h}$ and decodes to the input $\tilde{a} = \check{\chi}(h;\theta_D)$, where $\theta_E$ and $\theta_D$ are the encoder and decoder weights respectively. Before the autoencoder, we subtract the mean of each PCA mode and normalize by the largest standard deviation, $\bar{a}_i = (a_i - |a_i|)/\textrm{max(std(}a_i))$. The autoencoder minimizes the mean squared error (MSE) of reconstruction
\begin{equation}
    \mathcal{L}_{AE} = ||\bar{a} - \bar{\tilde{a}}||^2_2
\end{equation}
by learning the encoder and decoder weights via stochastic gradient descent as implemented in Keras \citep{chollet2015keras}. We vary the latent space dimension in the range $d_h = 3-20$ \citep{linot2020deep}. The MSE decreases with $d_h$ until a plateau value $d_h \approx 5$, which we select as the latent dimension. In Fig. \ref{fig:autoencoder} we show the dimension reduction process for a sample FFoRM streamline (see Fig. \ref{fig:fform_viz} for flow kinematics). We also visualize the leading 4 PCA modes. 

\subsection{Time evolution of microstructure}
Next we learn the time evolution of the reduced microstructural variable $h$ and the corotating phase under flow as
\begin{equation}
\begin{split}
    \frac{dh}{d\lagtime} &=  g_1(h,\sin\alpha,\cos\alpha,\textbf{Q}^{\alpha T} \cdot \textbf{Q}^T \cdot \textbf{E} \cdot \textbf{Q}^{\alpha} \cdot \textbf{Q};\theta_{g1}) \\
    \frac{d\alpha}{d\lagtime} &=  g_2(h,\sin\alpha,\cos\alpha, \textbf{Q}^T \cdot \textbf{E} \cdot \textbf{Q};\theta_{g2})
\end{split}
\label{eq:evolution}
\end{equation}
where $g_1$ and $g_2$ are neural network representations of the vector fields governing the latent space and phase evolutions respectively. \MDGrevise{An evolution equation with this structure is sometimes referred to as a} neural ODE (NODE) \citep{chen2018neural}. 
We use two NNs because the autoencoder latent space $h$ evolves in the phase-aligned corotating frame, whereas the phase evolves in the corotating frame. We input only the rate of strain tensor rather than the velocity gradient tensor because rotation of fluid elements is accounted for by the corotating reference frame. Our choice of inputs for the corotating phase evolution is physically informed by the evolution equation of a material line in the corotating frame \citep{Graham.2018.https://doi.org/10.1017/9781139175876}\MDGrevise{, as we illustrate in the Appendix}. Similarly, we transform the rate of strain tensor from the corotating to the phase-aligned frame by $\textbf{Q}^\alpha$ for pattern evolution (see Appendix for details). 

Finally, we input $\sin \alpha$ and $\cos \alpha$ rather than $\alpha$ to allow for continuous phase evolution. The phase is bounded by $\alpha \in [-\pi,\pi]$, but there is a discontinuous change when $\alpha$ passes over the boundary at $\pm \pi$. Whenever this occurs in the training data, we add or subtract $\pi$. This resolves the discontuity and the resulting rotation is the same, but it also causes a problem at long times when the phase can become arbitrarily large. The predicted $\alpha$ is then outside the range of the training data, and NNs perform significantly worse. Using $\sin \alpha$ and $\cos \alpha$ maps these values back to the range $[-1,1]$, allowing time evolution for unlimited duration.

To train the NODEs, we use a Runge-Kutta 45 integrator to forecast the latent space and phase
\begin{equation}
\begin{split}
    \hat{h}(\lagtime + \tau) &= h(\lagtime) + \int_\lagtime^{\lagtime+\tau} g_1(h,\sin \alpha,\cos\alpha, \textbf{Q}^{\alpha T} \textbf{Q}^T \cdot \textbf{E} \cdot \textbf{Q} \cdot \textbf{Q}^{\alpha};\theta_{g1}) d\tau
    \\
    \hat{\alpha}(\lagtime + \tau) &= \alpha(\lagtime) + \int_\lagtime^{\lagtime+\tau} g_2(h,\sin \alpha,\cos\alpha, \textbf{Q}^T \cdot \textbf{E} \cdot \textbf{Q} \cdot;\theta_{g2}) d\tau
\end{split}
\end{equation}
The model is trained to minimize the loss $\mathcal{L}_g = ||h(\lagtime + \tau) - \hat{h}(\lagtime + \tau)||^2_2 + ||\alpha(\lagtime + \tau) - \hat{\alpha}(\lagtime + \tau)||^2_2$ by stochastic gradient descent using automatic differentation to perform backpropagation \citep{chen2018neural, linot2022data}. We use the training forecast horizon $\tau = 0.1/D_r$, where $D_r$ is the rod rotational diffusion constant. Finally, predictions can be decoded to the full state scattering intensity $\tilde{\hat{I}}_{\bm{\Omega},\alpha}(\bm{q},\lagtime) = \check{\chi}(\hat{h}(\lagtime))U^T$ and rotated to the laboratory frame by the predicted phase and the vorticity tensor as $\tilde{\hat{I}}(\bm{q},\lagtime) = \mathcal{R}(\mathcal{R}(\tilde{\hat{I}}_{\bm{\Omega},\alpha}(\bm{q},\lagtime),-\hat{\alpha}(\lagtime)),-\theta_{\bm{\Omega}}(\lagtime))$.

\subsection{Stress}
Given stress data, an additional neural network can be trained to map the latent representation of the microstructure to the phase-aligned stress field: 
\begin{equation}
    \tilde{\bm{\sigma}}_p^{\alpha} = F(h,\textbf{Q}^{\alpha T} \cdot \textbf{Q}^T \cdot \textbf{E} \cdot \textbf{Q} \cdot \textbf{Q}^{\alpha};\theta_F)\label{eq:stressrot}
\end{equation}
We subtract the mean from each stress component and normalize the largest standard deviation $\bar{\sigma}_{p,ij}^\alpha = (\sigma_{p,ij}^\alpha - |\sigma_{p,ij}^\alpha|)/ \textrm{max(std(}\sigma_{p,ij}^\alpha))$ and use stochastic gradient descent to minimize the loss $L_F = ||\bar{\bm{\sigma}}_p^{\alpha} - \bar{\tilde{\bm{\sigma}}}_p^{\alpha}||^2_2$. The phase-aligned stress can then be transformed back to the lab frame by $\bm{\sigma}_p = \textbf{Q}^{\alpha} \cdot \textbf{Q} \cdot \bm{\sigma}^{\alpha}_p \cdot \textbf{Q}^T \cdot \textbf{Q}^{\alpha T}$. This function and Eq.~\ref{eq:evolution} for the microstructure gives the evolution of the stress -- a tensorial frame-indifferent constitutive model determined from FFoRM-sSAXS data.

We will not tackle the problem of inferring stress from velocity measurements in the present work. Nevertheless,  we provide here a sketch of how this might be done. For a solution of particles or polymers in a Newtonian solvent with viscosity $\eta_s$, and neglecting inertia, the momentum equation is
 $\divg\bsigma_p=-\eta_s\nabla^2\bm{v}$, where $\bsigma_p(\bm{x})$ is the desired polymer stress (inertia can be included if necessary) \citep{batchelor1970stress}. In an incompressible fluid the isotropic part of the stress is indeterminate, so we can take $\bsigma_p$ to be traceless. To deal with noise, we can use the fact that the flow is steady so we can use velocimetry data from multiple time instants. To determine $\bsigma_p$ from measurements of $\bm{v}$, a variational data assimilation framework, which requires no derivatives of velocity data, can be used \citep{Asch:2016}. Here we find the solution (stress and velocity) to $\divg\bsigma_p=-\eta_s\nabla^2\bm{v}$ and $\divg\bm{v}=0$ that minimizes $\mathbb{E}\{ ||\bm{v}-\bm{v}_{e}||^2\}$, where $\bm{v}_e$ is the experimental velocity, and $\mathbb{E}\{\cdot\}$ is expected value over the data. Alternately, physics-informed neural networks \citep{Karniadakis.2021.10.1038/s42254-021-00314-5,Thakur.2022.10.48550/arxiv.2209.06972} have been recently advocated for solving related problems --  this approach may also be useful for the present problem. 
 
\subsection{Synthetic training data \label{subsec:data}}

In the present work we develop and validate SIMPLE on a synthetic data set intended to represent the dispersions of rigid, rod-like cellulose nanocrystals (CNCs) used in the FFoRM-sSAXS experiments of \cite{corona_fingerprinting_2022}, although simplifications have been made. For further details of the device geometry, we refer to this previous work. We first generate Lagrangian flow trajectories in a model FFoRM by OpenFOAM simulations as performed in \cite{corona_probing_2018} assuming the fluid is Newtonian. We vary the ratio of the inlet flow rate $Q_1$ and outlet flow rates $Q_2$ over a range $Q_1 / Q_2 = (-0.7,-0.6,0.5,...,1.0)$, yielding flows that vary from shear to extension at the stagnation point as characterized by the nominal flow type parameter
\begin{equation}
    \Lambda = \frac{|\textbf{E}| - |\bm{\Omega}|}{|\textbf{E}| + |\bm{\Omega}|}
\end{equation}
where $|\bm{A}| = (\bm{A}:\bm{A}^T)^{1/2}$. The velocity gradient tensor for a planar flow can be expressed in terms of the flow type parameter and flow strength $G = |\bm{\nabla v}|/\sqrt{1+\Lambda^2}$ with cartesian axes coincident with the principle rate of strain axes
\begin{equation}
    \bm{\nabla v} = \frac{G}{2} \begin{bmatrix} 1 + \Lambda & 1 - \Lambda & 0 \\ -(1 - \Lambda) & -(1 + \Lambda) & 0 \\ 0 & 0 & 0 \end{bmatrix}
    \label{eqn:gradv}
\end{equation}
Streamlines are collected in the central stagnation region of the device, and velocity gradient components in the $z$-plane are set to 0. In experiments the scattering intensity is averaged over the depth of the device, although for sufficient channel height these effects are small \citep{corona_probing_2018}.

We vary the inlet flow rate $Q_1 \approx 0.025 - 0.5 $mL/min, corresponding to a dimensionless rotational Peclet number $\textrm{Pe} = |\textbf{E}|/D_r \approx 1 - 20$ for the CNC suspensions, where $D_r = 2 s^{-1}$ is the rotational diffusion constant, and $\eta_s \approx 1 $ Pa s is the solvent viscosity. We nondimensionalize the velocity gradient and time scales from OpenFOAM simulations as $\bm{\nabla v}^* = \bm{\nabla v}/D_r$ and $t^* = tD_r$. For notational convenience the $*$ superscripts are dropped hereafter. For each combination of flow rate and inlet/outlet ratio we sample 20 streamlines passing through the crossflow region, yielding $\sim 2000$ streamlines in total. Compared to the resolution of streamlines available in the experimental FFoRM-sSAXS data set, this represents a significant undersampling. Future work will investigate the influence of the range of flow type, flow strength, and size of the training data upon SIMPLE performance.

The velocity gradient tensor is then used to perform Brownian dynamics simulations of dilute, high-aspect ratio cylindrical rods as in \cite{tao2005brownian, tao2005kayaking}. We simulate $10^5$ trajectories to provide nearly noise-free measurements of the second and fourth orientation tensor moments, $\langle \textbf{uu} \rangle$ and $\langle \textbf{uuuu} \rangle$, the OPDF in spherical coordinates $N(\theta_{rod},\phi_{rod})$, and the scattering intensity $I(\bm{q})$. 

The scattering intensity calculation assumes a dilute cylindrical rigid rod suspension
\begin{equation}
    \overline{I(\bm{q})} = \phi (\Delta \rho)^2 V_p \overline{P(\bm{q}) S(\bm{q})} + b
\end{equation}
Overbars indicate an ensemble average over the BD trajectories, or equivalently over time for a steady state flow. For a dilute suspension interparticle scattering effects are insignificant, $\overline{P(\bm{q}) S(\bm{q})} \approx P(\bm{q})$. The model parameters are selected based on a fit to the CNCs and solvent in \cite{corona_fingerprinting_2022}, yielding the particle length $L = 1680$ nm, radius $R = 34$ nm, particle volume fraction $\phi = 0.1$, scattering length density difference between particle and solvent $\Delta \rho = 3.2 \times 10^{-6}$ \r{A}$^2$, particle volume $V_p = \pi R^2 L$, and incoherent scattering cross-section $b = 1\times10^{-4}$ cm$^{-1}$. For calculation of the form factor we refer to \cite{corona_bayesian_2021}, which integrates the single-particle scattering over the OPDF, $N(\theta_{rod},\phi_{rod})$. The particle contribution to the stress is calculated from the orientation tensor moments using Batchelor's expression for noninteracting particles \citep{batchelor1970stress, corona_bayesian_2021}. Finally, all quantities are sampled every $0.01$ dimensionless time units, yielding $\approx 2 \times 10^5$ total data points. Of these, half of the flow rates are retained for test data, so the total training data size is $M \approx 1 \times 10^5$.

\section{Results and Discussion \label{sec:results}}
We now evaluate SIMPLE on the data described above, starting with training of data-driven operations and interpolation within the training data. We then extrapolate the model to startup steady shear and extensional flow protocols.
\subsection{Training and Interpolation \label{subsec:training}}
In Table \ref{table:loss} we report the error of data-driven operations in SIMPLE evaluated on 720 FFoRM streamlines. These data are within the range of flow rates $Q_1 + Q_2$ and flow types $\Lambda$ used in training, but were not explicitly used to train the NNs or perform the singular value decomposition. We report the normalized reconstruction error of the phase-aligned scattering intensity from the truncated PCA basis, $|I_{\bm{\Omega},\alpha} - \tilde{I}_{\bm{\Omega},\alpha}|/|I_{\bm{\Omega},\alpha}|$, as well as the NN losses for the autoencoder, NODE, and the stress function. The NN architectures are reported in the Appendix.

\begin{table}
\setlength{\tabcolsep}{12pt}
\centering
\begin{tabular}{c c c c}
\hline \hline
$|I_{\bm{\Omega},\alpha} - \tilde{I}_{\bm{\Omega},\alpha}|/|I_{\bm{\Omega},\alpha}|$ & $\mathcal{L}_{AE}$ & $\mathcal{L}_g$ & $\mathcal{L}_F$ \\
\hline
$8.2 \times 10^{-4}$ & $1.2 \times 10^{-6}$ & $9.1 \times 10^{-4}$ & $7.1 \times 10^{-4}$ \\
\hline
\end{tabular}
\caption{Test error of data-driven operations.}
\label{table:loss}
\end{table}

\begin{figure*}
    \centering
    \includegraphics[width=1.0\textwidth]{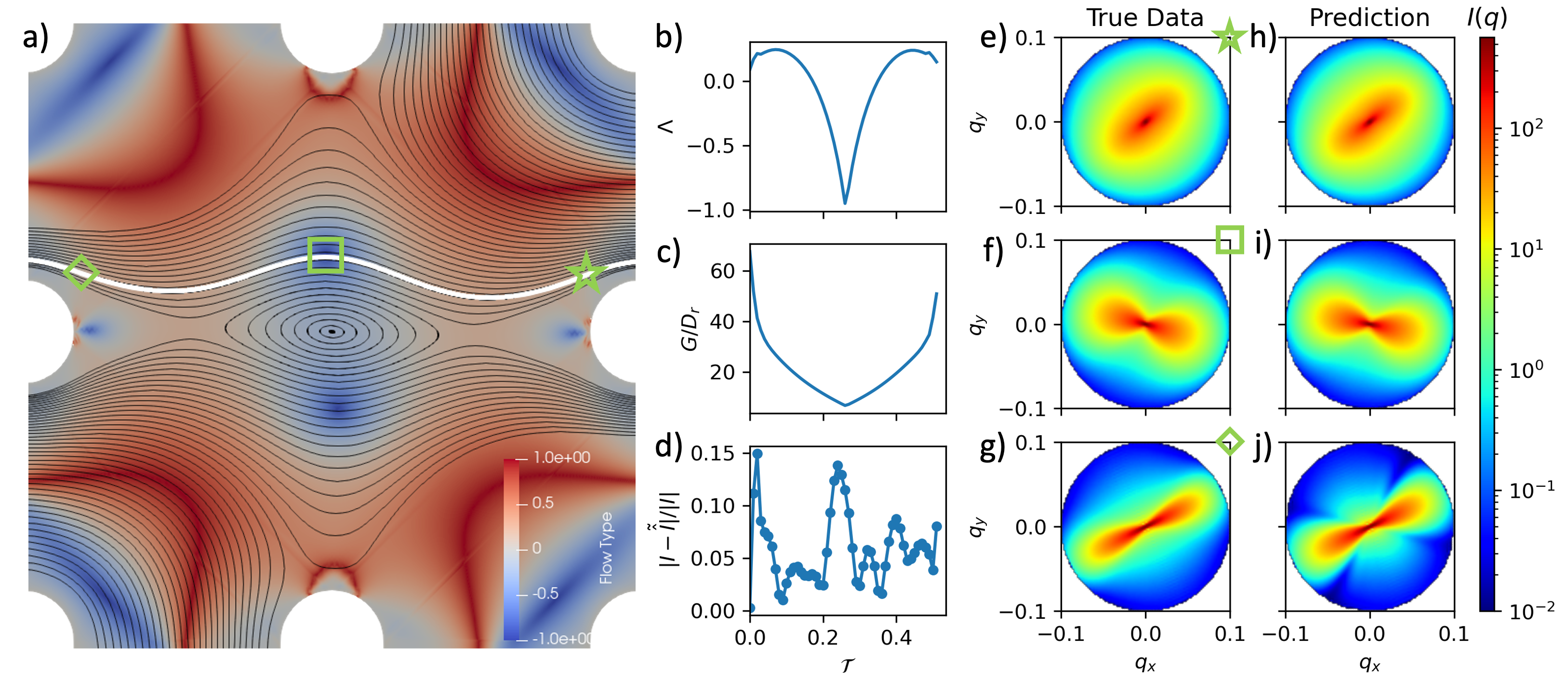}
    \caption{Microstructural evolution prediction along a Lagrangian streamline in the FFoRM. a) Streamlines in the FFoRM in black (white line is the selected streamline) and snapshot locations of the sample predictions in (h-j) (green symbols) b) Flow type parameter $\Lambda$ c) Flow strength $G$ d) Prediction error versus Lagrangian time $\lagtime$. e-g) True data from BD simulations. h-j) SIMPLE predictions evolved from initial condition $I(\lagtime = 0)$ with input $\bm{\nabla v}(\lagtime)$ as described in the text. Snapshot time $\lagtime = 0.01$ (star), $\lagtime = 0.24$ (square), and $\lagtime = 0.52$ (diamond).}
    \label{fig:fform_viz}
\end{figure*}

As a visual example, we demonstrate the predictive capabilities of the trained SIMPLE model for a specific streamline from the FFoRM test data. In particular, we choose the operating conditions $Q_1/Q_2 = - 0.7$ and $\textrm{Pe} \approx 20$, in which case the flow field exhibits significant spatial fluctuations and large Lagrangian excursions in $\Lambda$ and $G$. We time evolve the microstructure from $I(\lagtime = 0)$ corresponding to the start of the beam region (Fig. \ref{fig:fform_viz}a, star symbol) for a specified deformation $\bm{\nabla v}(\lagtime)$. Specifically, we rotate the initial condition to the phase-aligned reference frame $I_{\bm{\Omega},\alpha}(\lagtime=0) = \mathcal{R}(\mathcal{R}(I(\lagtime=0),\theta_{\bm{\Omega}}(\lagtime = 0)),\alpha(\lagtime=0))$, project onto the truncated PCA basis, and encode to the latent space $h(\lagtime=0) = \chi(\mathcal{P}_{d_a} U I_{\bm{\Omega},\alpha}(\lagtime=0);\theta_E)$. We then use the trained NNs in Eq. \ref{eq:evolution} to time evolve the latent space and phase, yielding $\hat{h}(\lagtime)$ and $\hat{\alpha}(\lagtime)$. We decode the latent space predictions to the scattering intensity $\tilde{\hat{I}}_{\bm{\Omega},\alpha}(\lagtime) = \check{\chi}(\hat{h}(\lagtime);\theta_D)U^T$, and finally rotate to the lab frame using the predicted phase $\tilde{\hat{I}}(\lagtime) = \mathcal{R}(\mathcal{R}(\tilde{\hat{I}}_{\bm{\Omega},\alpha}(\lagtime),-\hat{\alpha}(\lagtime)),-\theta_{\bm{\Omega}}(\lagtime))$.

We find the SIMPLE model tracks the scattering pattern and phase successfully, and the autoencoder decodes to the full space with low error -- typically several percent. The error is largest at the inlet due to the high flow strength $\textrm{Pe} \approx 15-20$. These are among strongest flows in the training data, which are the most difficult to predict, as we discuss further below. 
\MDGrevise{This error might be reduced by choosing the initial condition for the microstructure to be the steady state at the inlet shear rate.}
The error increases again at $\lagtime \approx 0.25$, where the scattering intensity undergoes a rapid reorientation due to the change from shear to rotational flow and back to shear flow. The primary NODE error comes in predicting the phase, but after this large flow gradient, the NODE tracking successfully reproduces the true trajectory again. SIMPLE error in predicting the stress from the time evolved latent space for this streamline is also low, as discussed in Sec. \ref{sec:rheology} (Fig. \ref{fig:interp_stress}).

\begin{figure}
    \centering
    \includegraphics[width=0.5\textwidth]{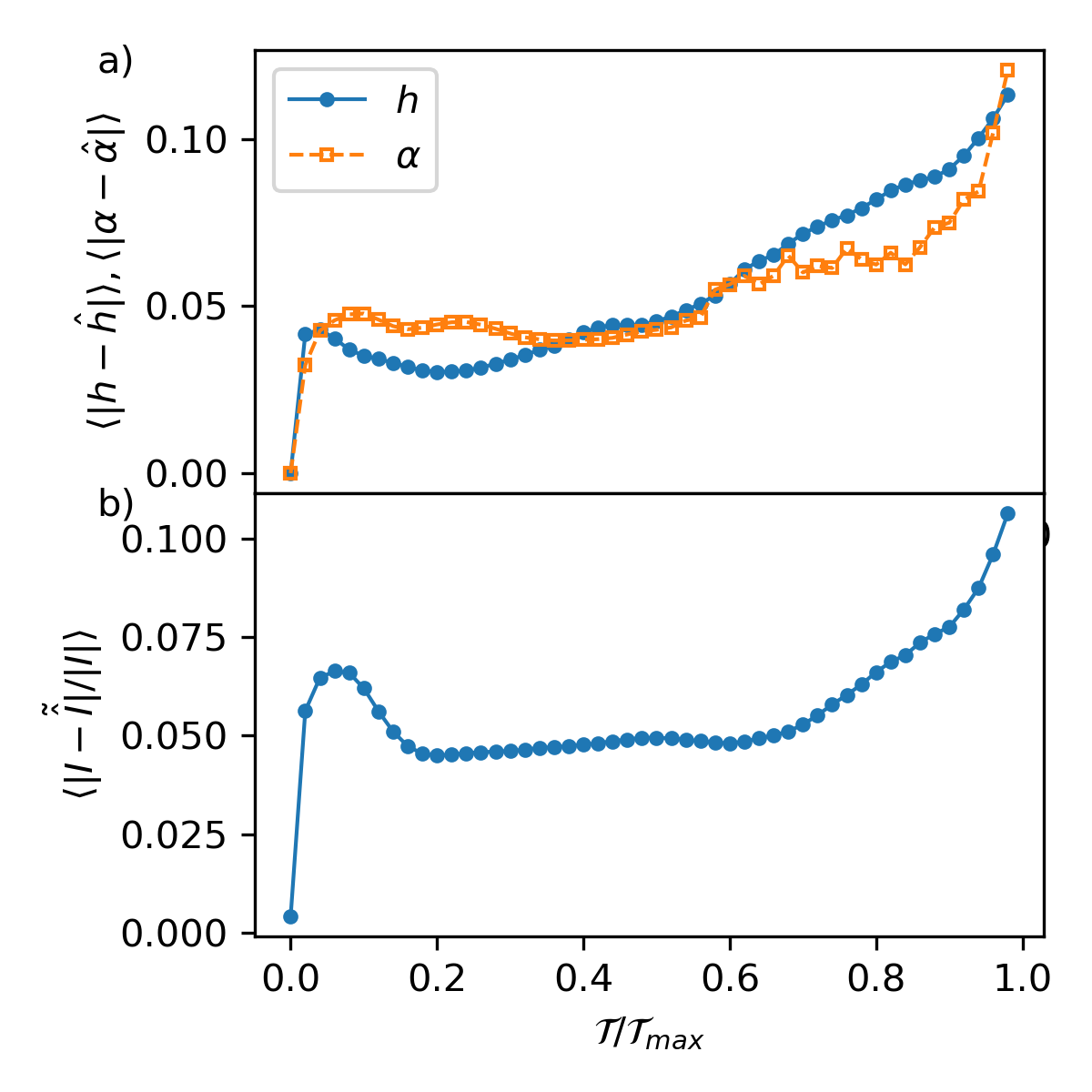}
    \caption{Ensemble average error of SIMPLE predictions evaluated on test FFoRM streamlines as a function of residence time a) Forecast error of the latent variables and phase predicted by the NODE b) Combined error of forecasting, decoding, and rotating to the lab frame.}
    \label{fig:interp}
\end{figure}

Next, we quantify ensemble error of predictions on test FFoRM streamlines. Generally the streamline time $\lagtime_{max}$ varies with the starting position $\textbf{X}_0$ and operating conditions, so we normalize each streamline time by its maximum duration, $\lagtime/\lagtime_{max}$, and report the ensemble average error. We show the error in the prediction of the latent space variables $h$ and co-rotating phase $\alpha$ in Fig. \ref{fig:interp}a. We also show the error in the lab frame scattering intensity in Fig. \ref{fig:interp}b after decoding the latent space forecast to the phase-aligned intensity and rotating back to the lab frame using the predicted phase. The forecasting error grows quickly at $\lagtime/\lagtime_{max} \approx 0$ and $\lagtime/\lagtime_{max} \approx 1$ due to large flow gradients at the beginning and end of the stagnation region. In the center of the device near the stagnation point the flow is relatively homogeneous, and the error is constant.

\subsection{Extrapolation \label{subsec:extrapolation}}
We have demonstrated the SIMPLE approach successfully predicts microstructural evolution under diverse deformation histories for FFoRM test data. A greater challenge is to extend model predictions to deformation inputs and histories that are substantially different than those used for training. We will demonstrate the extrapolation capabilities of the SIMPLE approach in three ways: 1) startup of steady shear flow, which is not explicitly in the FFoRM data; 2) startup of steady planar extensional flow along an extension axis not seen in the FFoRM; 3) Varying the flow strength $\textrm{Pe}$ beyond the training data range. 

We generate scattering intensity patterns from BD simulations using a constant velocity gradient tensor given by Eq. \ref{eqn:gradv} with $\Lambda = 0$ for shear flow and $\Lambda = 1$ for extensional flow. The initial condition is an isotropic suspension. In the FFoRM training data set, there are no streamlines in which the fluid undergoes simple shear deformation for an extended duration. There are streamlines comparable to startup to startup planar extensional flow, as seen in Fig. \ref{fig:Fig2}b,c (left column), but the axis of extension is at an angle of $45^\circ$ with respect to the $x$-axis, as compared to the current case where the extension is $x$-axis aligned. We vary the flow strength $G/D_r = 0.1-50.0$, as compared to the training data range $\textrm{Pe} \approx 1-20$.

\begin{figure*}
    \centering
    \includegraphics[width=1.0\textwidth]{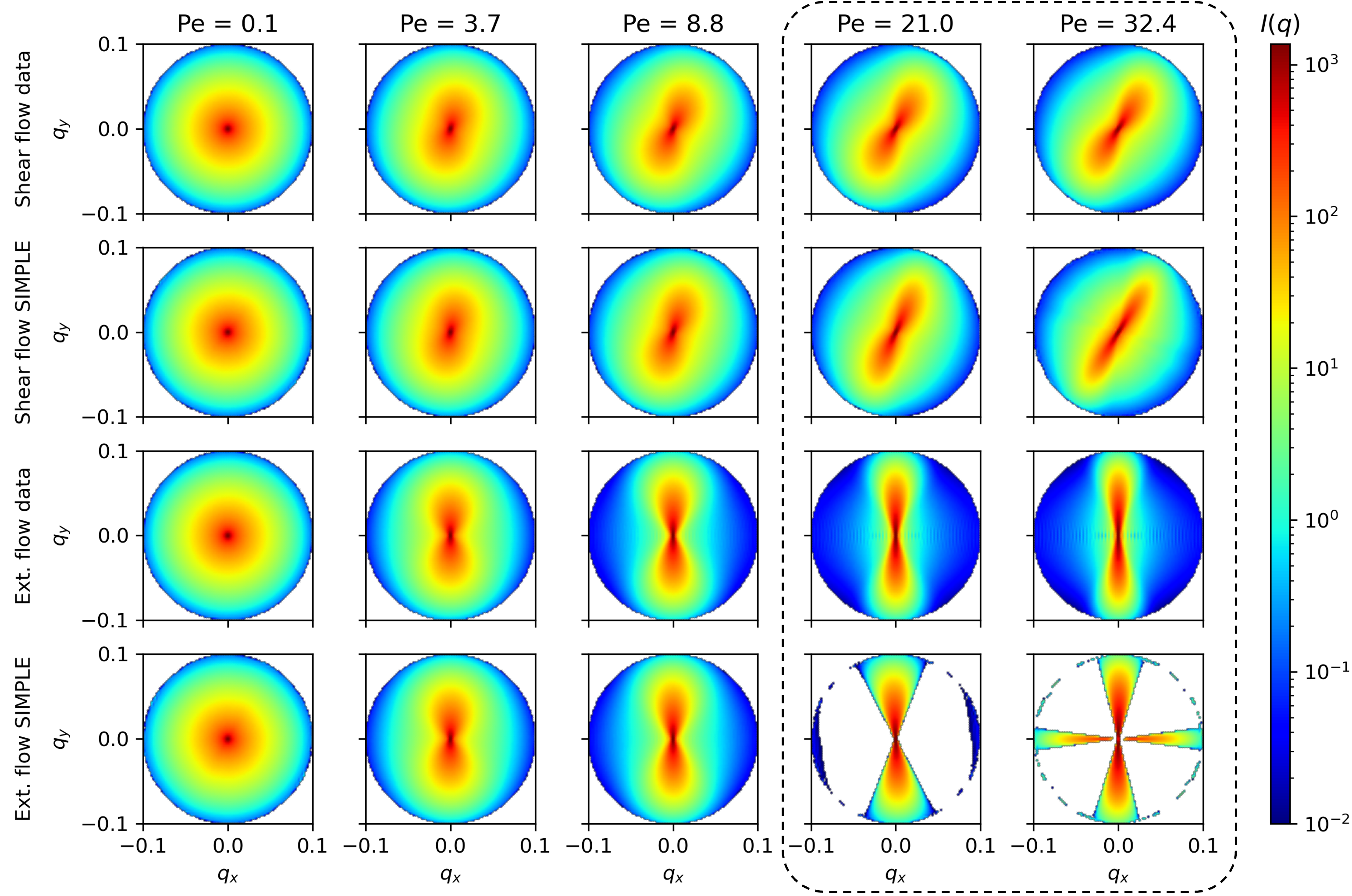}
    \caption{Steady state scattering intensity from BD simulations and SIMPLE predictions under shear and extensional flow for increasing $\textrm{Pe} = \dot{\gamma}/D_r$ and $Pe = \dot{\epsilon}/D_r$ respectively from left to right. First row: BD simulation shear flow. Second Row: SIMPLE prediction shear flow. Third row: BD simulation extensional flow. Fourth row: SIMPLE prediction extensional flow. The dashed box indicates flow conditions which are above the training data range $\textrm{Pe} = 1-20$.\MDGrevise{The white regions on the rightmost two scattering patterns on the bottom row indicate spurious negative intensity predictions.}}
    \label{fig:ss_snaps}
\end{figure*}

Using the same velocity gradient tensor $\bm{\nabla v}(\lagtime)$ as the BD simulations, we make SIMPLE predictions from an initially isotropic scattering intensity as detailed in Sec. \ref{subsec:training}. We qualitatively visualize these predictions in comparison the the BD simulations in Fig. \ref{fig:ss_snaps}. Generally, SIMPLE accurately predicts the qualitative scattering pattern and long-time steady state behavior. For increasing flow strength, the prediction error grows. We primarily attribute this to the range of training data, $\textrm{Pe} \approx 1-20$. For a purely data-driven framework, we expect the model to perform poorly where it has not been trained. This could be addressed by physically informed approaches to time evolution \citep{mahmoudabadbozchelou_nn-pinns_2022}, or by multi-fidelity modeling where high Pe data from an imperfect model is used to supplement the FFoRM data \citep{Mahmoudabadbozchelou:2021hk, Mahmoudabadbozchelou:2021bf}. The white regions for SIMPLE predictions in extensional and $\textrm{Pe} > 20$ correspond to negative values of scattering intensity, which is clearly an unphysical result. These emerge from errors in the reconstructed predictions of the PCA modes, which can be negative valued. However, this only occurs for scattering patterns outside the training data range. This could be addressed by constraining the reconstruction of PCA modes to be positive valued.

\begin{figure*}
    \centering
    \includegraphics[width=1.0\textwidth]{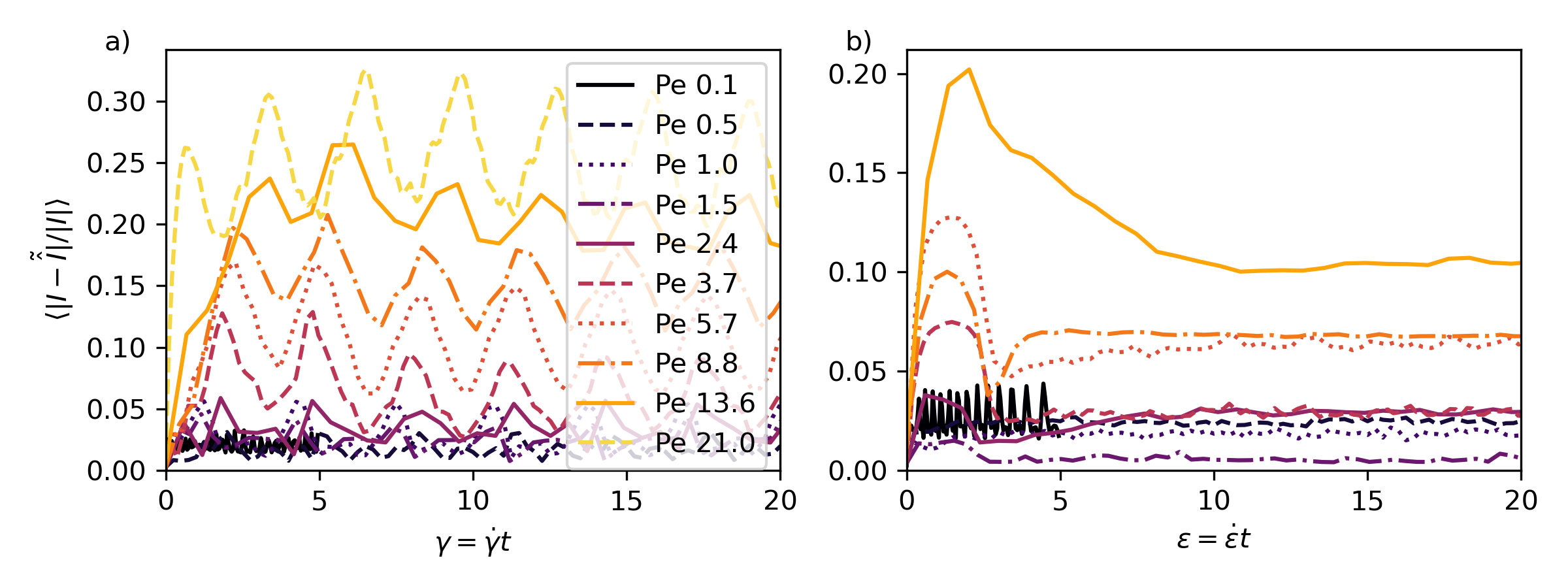}
    \caption{SIMPLE prediction error versus accumulated strain for increasing flow strength under startup of a) shear flow b) extensional flow.}
    \label{fig:extrap_rmse}
\end{figure*}

We quantify the prediction error in these extrapolation flow protocols in Fig. \ref{fig:extrap_rmse}. We observe oscillations in the shear flow predictions and the extensional flow predictions at low Pe. These emerge due to a lag between the predicted and true phase $\alpha$ (Appendix Fig. \ref{fig:phase_lag}). Because the phase NN evolves under a corotational rate of strain tensor input, any phase lag in a constant vorticity flow will emerge as an oscillation, which is then propagated through the latent space $h$ prediction via the rotation of the rate of strain tensor to the phase-aligned reference frame. Additionally, the equilibrium phase is undefined, which causes challenges in data-driven modeling \citep{linot2020deep} and leads to the low-Pe oscillations seen in the extensional flow predictions. In future work, we plan to address this issue by physically informed NNs, with the form $d \alpha/d\lagtime = f(\bm{\Omega}(\lagtime)) + g_2(h,\sin \alpha, \cos \alpha, \textbf{E};\theta_{g2})$, where the first term is known via the co-rotating frame evolution with the vorticity tensor.

\subsection{Rheology predictions \label{sec:rheology}}

\begin{figure}
    \centering
    \includegraphics[width=0.5\textwidth]{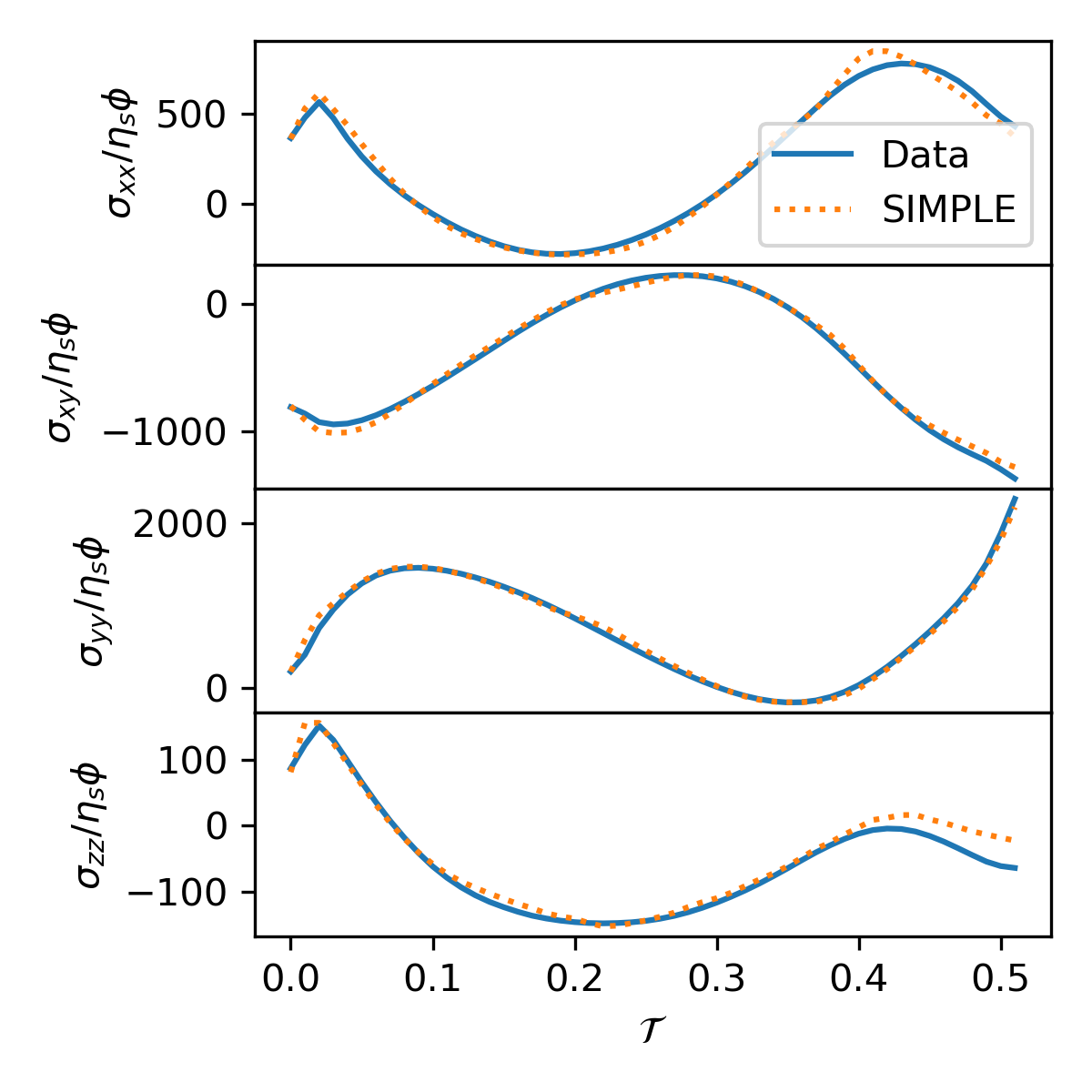}
    \caption{SIMPLE predictions of the stress for the FFoRM test streamline shown in Fig. \ref{fig:fform_viz}.}
    \label{fig:interp_stress}
\end{figure}

\begin{figure}
    \centering
    \includegraphics[width=0.5\textwidth]{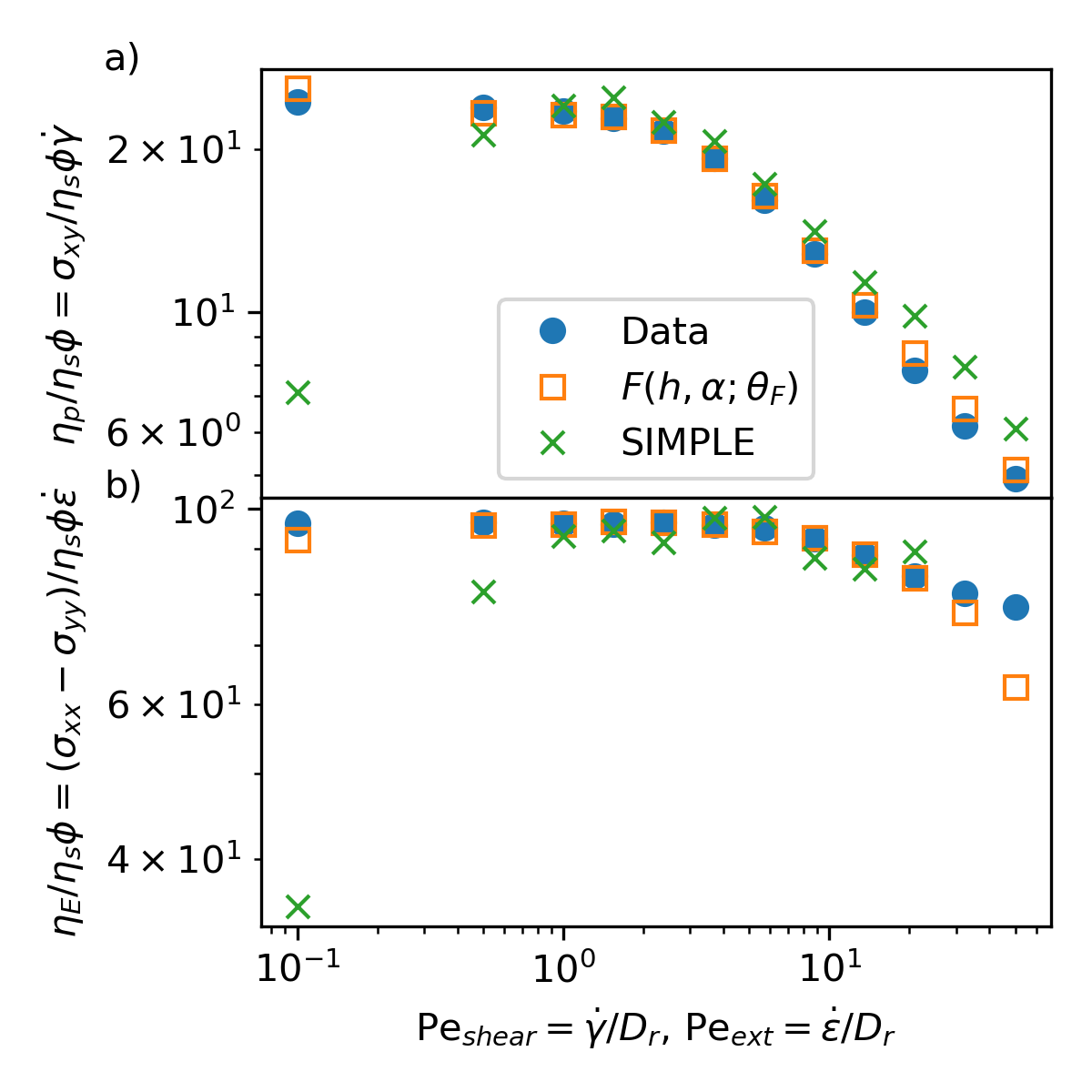}
    \caption{BD simulation results (blue circles), SIMPLE predictions (green x's), and stress function predictions given the true scattering intensity $F(h,\alpha)$ (orange squares) for the particle contribution to the viscosity in a) steady shear flow b) steady extensional flow.}
    \label{fig:extrap_stress}
\end{figure}

Finally, we demonstrate the ability of NNs to learn the particle stress from the latent space scattering data in Fig. \ref{fig:extrap_stress}. As discussed above, we consider an ideal case in which training data for the stress is available from BD simulations. To obtain stress predictions, we take the time evolved latent space and phase for a given initial condition and deformation (as described in Sec. \ref{subsec:training}) and input them to the stress function $\tilde{\hat{\bm{\sigma}}}_p^\alpha = F(\hat{h},\textbf{Q}^{\hat{\alpha} T} \cdot \textbf{Q}^T \cdot \textbf{E} \cdot \textbf{Q} \cdot \textbf{Q}^{\hat{\alpha}};\theta_F)$. In Fig. \ref{fig:interp_stress} we consider SIMPLE predictions for the FFoRM test streamline shown in Fig. \ref{fig:fform_viz}. We find SIMPLE reproduces the true stress accurately, even at $\lagtime \approx 0.25$ where the scattering intensity prediction error is relatively high. 

We then predict the stress in steady shear and extensional flow using the SIMPLE results from Sec. \ref{subsec:extrapolation}. We exclude the first 10 strain units of data before computing the steady state shear viscosity and extensional viscosity. In Fig. \ref{fig:extrap_stress} we compare these predictions (green x's) to BD simulation results (blue circles). The model accurately predicts the shear viscosity within the training data range $\textrm{Pe} = 1-20$. With respect to the poor prediction at $\textrm{Pe} = 0.1$, calculating low Pe shear viscosity is challenging even from BD simulations \MDGrevise{because the signal-to-noise ratio is unfavorable in that regime.} SIMPLE predicts the extensional viscosity accurately up to $\textrm{Pe} \approx 20$. For stronger flows, SIMPLE predicts a negative extensional viscosity due to time evolution errors which lead to spurious alignment in the direction of compression, which is clearly visible in the bottom right panel of Fig. \ref{fig:ss_snaps}. \MDGrevise{Methods to combine data-driven modeling with asymptotic approaches for limiting cases such as $\textrm{Pe}\rightarrow 0$ and $\textrm{Pe}\rightarrow \infty$ will be the topic of future work.} 

To evaluate the accuracy of the stress function without time evolution errors, we also consider predictions of the stress given the true scattering intensity pattern (orange squares in Fig. \ref{fig:extrap_stress}). We rotate the true scattering intensity to the phase-aligned reference frame $I_{\bm{\Omega},\alpha}(\lagtime) = \mathcal{R}(\mathcal{R}(I(\lagtime),\theta_{\bm{\Omega}}(\lagtime)),\alpha(\lagtime))$, encode to the latent space $h(\lagtime) = \chi(\mathcal{P}_{d_a} U I_{\bm{\Omega},\alpha}(\lagtime);\theta_E)$, and make stress predictions $\tilde{\bm{\sigma}}_p^\alpha = F(h,\textbf{Q}^{\alpha T} \cdot \textbf{Q}^T \cdot \textbf{E} \cdot \textbf{Q} \cdot \textbf{Q}^{\alpha};\theta_F)$. We find the stress function accurately predicts the shear and extensional viscosity at low Pe, suggesting that errors in the full SIMPLE model arise primarily from \MDGrevise{the time evolution aspect of the model, not the dimension reduction}. 

\section{Conclusions}

The \Formsax~experiment originally demonstrated by \cite{corona_fingerprinting_2022} provides remarkable access to microstructural information in a flowing complex fluid.  
The present work demonstrates how to use this information in Lagrangian reference frames attached to a large number of streamlines in the flow to develop efficient and accurate low-dimensional models for the evolution of the microstructure.  
The approach we develop, called ``Scattering-Informed Microstructure Prediction during Lagrangian Evolution (SIMPLE)'', provides a data-driven workflow to rapidly develop general frame-indifferent evolution equations that can then be used to simulate that fluid in an arbitrary flow.
In this paper we have introduced the basic data-driven framework, using efficient data-augmentation methods to make the best use of the available data, as well as a corotating reference frame that enables the microstructural evolution model to obey the key physical principle of material frame indifference.  
Results thus far are very promising, both for ``interpolating" predictions for evolution along streamlines in the FFoRM that were not included in the training data, as well as for ``extrapolating" predictions for flow histories quite different than those found in the training data. In particular, from the Lagrangian data, we can reconstruct the correct evolution for startup of simple shear and planar extension.


Further methodological improvements to the SIMPLE frameworks are necessary. The small phase errors need to be reduced, and of course the larger issue of extracting stress from velocity fields must be addressed; we have described potential approaches for doing so. \MDGrevise{A potential concern regarding the current SIMPLE framework may be that the construction of the microstructural NN model is through reduced-order scattering information rather than direct microstructural descriptors, resulting in some lack of interpretability of the reduced description.  We note that accurate general approaches exist to convert scattering information directly to real-space information such as the OPDF and its moments (\cite{corona_bayesian_2021,corona2023testing}), so an alternative to the ML-based dimension reduction pursued here would be to explicitly specify physically-motivated descriptors to be used in the dynamic modeling aspect of the algorithm. This approach would likely trade efficiency for interpretability, since the ML-based approach uses an optimization procedure to determine the latent representation.}

More interestingly from the soft matter physics perspective, the SIMPLE framework can also be integrated with physics-based evolution equations for microstructure, rather than the data-driven ones considered here, whose only physical input is the constraint of frame indifference. For example, the SIMPLE formalism also provides an opportunity to combine physics- and data-driven modeling by advancing the development of closure models. For a suspension of rigid rods with orientation $\textbf{u}$, the evolution equation for the second moment of the OPDF, $\langle \textbf{uu}\rangle$, depends on the fourth moment $\langle \textbf{uuuu}\rangle$ and so on. No fully satisfactory closure model, even for this simplest complex fluid, exists -- for example none get the scaling for shear-thinning of viscosity correct \cite{corona2023testing}.  It may be possible to use data and symmetry, in a manner analogous to the turbulence closure work of Ling et al. \citep{Ling:2016fz} noted above, to find a closure model from data.  
For nondilute systems, a key challenge is to incorporate particle-particle interactions into the evolution equation for the orientational distribution function. In principle, a general theoretical framework for describing these effects already exists \citep{dhont2003viscoelasticity,lang2019microstructural}, but the functional forms of the interaction terms are not well understood. 
One potential approach is a hybrid of ML and data assimilation, where we take the form of the evolution equation to be known, but learn various functional contributions to it by representing them as neural networks (or some other functional representation) whose parameters are learned from the FFoRM-sSAXS data. More broadly, the most effective model in practice may be one that takes a physics-based model as a baseline, with a data-driven model learning the difference between the physics-based model and the data.

 Finally, real processing flows are almost never viscometric, whereas most rheological data come from viscometric flows. Therefore the general importance of a combined experimental and modeling framework that enables direct development of fully tensorial constitutive models from transient, nonviscometric data sets cannot be overstated. The combination of \Formsax~and SIMPLE provides an important step toward this aim. 

\section*{Declarations}
\subsection*{Funding}
MDG and CY acknowledge support from ONR N00014-18-1-2865 (Vannevar Bush Faculty Fellowship). \MEHrevise{MEH, PTC and AD acknowledge support from the Department of Energy, Award Number DE-SC0020988.}

\subsection*{Conflicts of interest/competing interests}
All authors certify that they have no affiliations with or involvement in any organization or entity with any financial interest or non-financial interest in the subject matter or materials discussed in this manuscript.

\section*{Appendix}\label{sec:appendix}
\subsection*{Neural network architectures}
The neural network architectures and activation functions are given in Table \ref{table:arch}. The depth and width were increased empirically until improvement in the test loss plateaued. All models are trained using the Adam optimizer in PyTorch. We use the PyTorch package torchdiffeq \citep{chen2018neural} to train the NODEs.

\begin{table*}[h]
\setlength{\tabcolsep}{12pt}
\centering
\begin{tabular}{c c c c}
\hline \hline
Function & Shape & Activation \\
\hline
$\chi$ & $d_a:64:128:64:d_h$ & ReLU:ReLU:ReLU:Linear \\
$\check{\chi}$ & $d_h:64:128:64:d_a$ & ReLU:ReLU:ReLU:Linear \\
$g_1$ & $d_h + 11:128:256:256:128:d_h$ & ReLU:ReLU:ReLU:ReLU:Linear \\
$g_2$ & $d_h + 11:32:64:32:1$ & ReLU:ReLU:ReLU:Linear \\
$F$ & $d_h + 9:128:128:9$ & ReLU:ReLU:Linear \\
\end{tabular}
\caption{NN architectures. The additional 11 inputs to the time evolution models $g_1$ and $g_2$ correspond to the rate of strain tensor $\textbf{E}$ and the phase $(\sin \alpha, \cos \alpha)$. The additional 9 inputs to the stress model $F$ correspond to the rate of strain tensor. In the current application to dilute rigid rod suspensions, $d_a = 50$, and $d_h = 5$.}
\label{table:arch}
\end{table*}

\subsection*{Evolution in the co-rotating frame}
In principle, the lab frame rate of strain tensor $\textbf{E}$ and the rotation matrix $\textbf{Q}$ should be sufficient for the SIMPLE to perform time evolution in the co-rotating frame. However, it improves accuracy and reduces the amount of necessary training data to input the rate of strain tensor in the appropriate reference frame, as we have done in Eqs.~\ref{eq:evolution}-\ref{eq:stressrot}. To illustrate the proper transformation of $\textbf{E}$ for the corotating frame, we consider here a simple case of evolution of an infinitesimal material line. In the Lagrangian, but non-rotating frame,
\begin{equation}
    \frac{d \textbf{u}}{d\mathcal{T}} = (\bm{\nabla v}^T) \cdot \textbf{u}
\end{equation}
The co-rotating frame orientation is related to the orientation in the nonrotating frame by $\textbf{u} = \textbf{Q} \cdot \textbf{u}^R$ where $\textbf{Q} = [\bm{e}_1^R,\bm{e}_2^R]$ is the unitary rotation matrix, which evolves according to Eq. \ref{eqn:vort_evo}, and the elements of $\textbf{u}^R$ are the components of $\textbf{u}$ in the corotating basis given by $\bm{e}_1^R$ and $\bm{e}_2^R$. This leads to
\begin{equation}
    \frac{d \textbf{u}}{d\mathcal{T}} = \frac{d \textbf{Q} \cdot \textbf{u}^R}{d\mathcal{T}} = \frac{d \textbf{Q}}{d\mathcal{T}} \cdot \textbf{u}^R + \textbf{Q} \cdot \frac{d \textbf{u}^R}{d\mathcal{T}} = (\textbf{E} + \bm{\Omega}^T) \cdot \textbf{Q} \cdot \textbf{u}^R
\end{equation}
Applying Eq. \ref{eqn:vort_evo}, left multiplying by $\textbf{Q}^T$, and noting $\textbf{Q}^T \cdot \textbf{Q} = \bm{\delta}$, \MDGrevise{the terms containing $\bm{\Omega}^T$ cancel out and }we obtain
\begin{equation}
    \frac{d \textbf{u}^R}{d\mathcal{T}} = \textbf{Q}^T \cdot \textbf{E} \cdot \textbf{Q} \cdot \textbf{u}^R
\end{equation}
Thus, $\textbf{Q}^T \cdot \textbf{E} \cdot \textbf{Q}$ \MDGrevise{is the appropriately transformed representation of the velocity gradient. }
Similarly, for the transform from the co-rotating frame to the phase-aligned frame we have
\begin{equation}
    \textbf{u}^R = \textbf{Q}^\alpha \cdot \textbf{u}^\alpha
\end{equation}
\begin{equation}
    \textbf{Q}^\alpha = \begin{bmatrix}
        \cos(\alpha) & -\sin(\alpha) \\ \sin(\alpha) & \cos(\alpha) \\
    \end{bmatrix}
\end{equation}
Following a similar process, after simplification we find
\begin{equation}
    \frac{d \textbf{u}^\alpha}{d\mathcal{T}} = \left( \textbf{Q}^{\alpha T} \cdot \textbf{Q}^T \cdot \textbf{E} \cdot \textbf{Q} \cdot \textbf{Q}^\alpha - \begin{bmatrix} 0 & -\alpha_t \\ \alpha_t & 0 \\ \end{bmatrix} \right) \cdot \textbf{u}^\alpha
\end{equation}
where $\alpha_t = d\alpha/d\mathcal{T}$. This informs the input to the NODE evolution of the phase-aligned scattering pattern.
\subsection*{Phase prediction lag}
\begin{figure}
    \centering
    \includegraphics[width=0.5\textwidth]{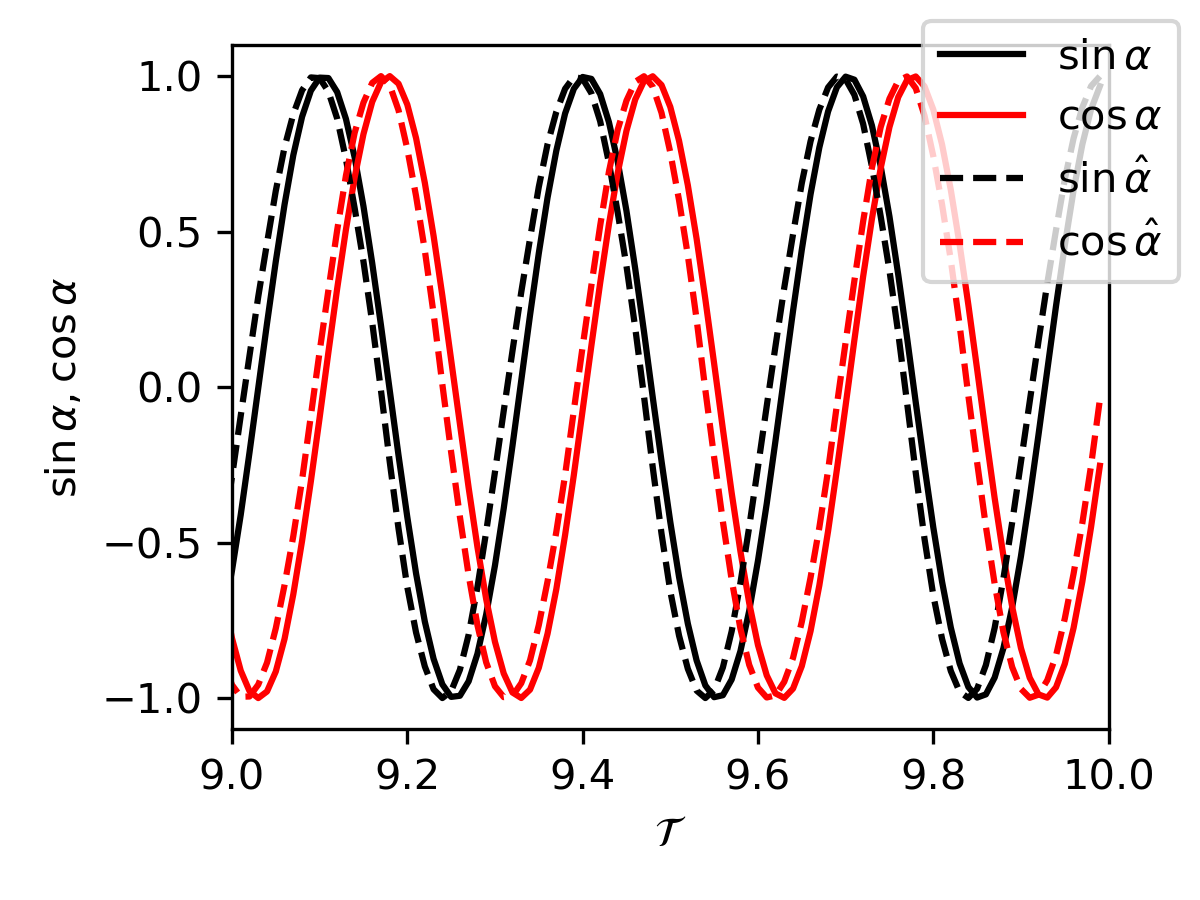}
    \caption{True and predicted $\sin \alpha$ and $\cos \alpha$ for steady shear flow $\textrm{Pe} = 21.0$.}
    \label{fig:phase_lag}
\end{figure}

In Fig. \ref{fig:phase_lag}, we show an example of the time evolution of the co-rotating phase in steady shear flow at $\textrm{Pe} = 21.0$. We observe a lag between the true and predicted phase which leads to the oscillations in error in Fig. \ref{fig:extrap_rmse}. This error emerges shortly after the start of prediction, and is constant in time thereafter. This is consistent across the range of $\textrm{Pe} = 1-50$.


\bibliography{bib_v1,MDGLibrary,MDGnewbib,MEHLibrary}

\end{document}